\documentstyle[aps,prl,preprint]{revtex}
\tightenlines
\include{psfig} 

\begin{document}

\draft \title{A semi-schematic model for the center of mass dynamics in
supercooled molecular liquids}

\author{Linda Fabbian, Francesco Sciortino, Filippo Thiery and Piero Tartaglia}
\address{Dipartimento di Fisica and Istituto Nazionale
 per la Fisica della Materia, Universit\'a di Roma {\it La Sapienza},
 P.le Aldo Moro 2, I-00185, Roma, Italy}

\date{\today}
\maketitle
\begin{abstract}
\noindent
We introduce a semi-schematic mode-coupling model to describe the slow
dynamics in molecular liquids, retaining explicitly only the
description of the center of mass degrees of freedom.  Angular degrees
of freedom are condensed in a $q$-vector independent coupling
parameter.  We compare the time and $q$-dependence of the density
fluctuation correlators with numerical data from a $250 ~ns$ long
molecular dynamics simulation.  Notwithstanding the choice of a
network-forming liquid as a model for comparing theory and simulation,
the model describes the main static and dynamic features of the
relaxation in a broad $q$-vector range.
\end{abstract}


%\begin{multicols}{2}

The slow dynamics in supercooled liquids has recently become 
a field of increasing interest\cite{review} and a large amount of 
knowledge has been accumulated from theoretical 
work\cite{review-glass}, experiments\cite{cummins,andalo} and 
simulations\cite{barrat,walstrom,kob}.  
It has now become apparent that the ideal mode-coupling theory (MCT)
for structural relaxation\cite{review-glass,Gotze}
contains the essential ingredients for the
description of the molecular dynamics in supercooled
states, down to a cross-over temperature $T_c$, where
``hopping'' processes become dominant\cite{extended,parisi}.\\
Although the MCT
was originally developed for simple (atomic) liquids,
it appears to be able to rationalize in a coherent way
experimental\cite{cummins,alba,sokolov} and numerical simulation 
results\cite{walstrom,kob-dumb,sgtc} 
also for
molecular liquids, {\it i.e.} for molecules with  anisotropic interaction
potentials.
The lack of theoretical predictions for
the behavior of the center of mass (COM) and 
angular degrees of freedom in molecular liquids  has been
compensated by the use of {\it schematic models}, where the
spectrum of different $q$-vectors and orientations is condensed into one
or two representative correlators. The exact solution of
these schematic
models and the comparison with experiments
\cite{alba,gotze-glicerolo} and simulations\cite{tei} provide
compelling evidence that even molecular liquids can be described 
in a MCT framework. A full and detailed study assessing 
the ability of MCT to predict not only the general features of 
supercooled molecular liquids, but also the $q$-dependence of the
COM correlators is highly desirable.\\ 
The possibility of describing theoretically the slow dynamics 
in supercooled molecular liquids is of fundamental 
importance for assessing the limitations of the ideal MCT, 
in making contact with experiments,
and in tackling the problem of the relation between translational and
rotational motions in supercooled states.
Theoretical work has recently been
undertaken\cite{schilling,theis,franosch-dumbell,kawasaki} 
in this direction and appears to offer 
in future the possibility of a detailed comparison 
between experiments and theory.\\
In this Letter we propose a {\it semi-schematic}\cite{schem} 
mode-coupling model to
describe the COM dynamics in molecular liquids and we compare the
predictions of the model with the COM collective correlation functions, 
calculated from a long molecular dynamics (MD) simulation for a model 
of a network forming molecular 
liquid\cite{sgtc}. For the first time, the comparison is
performed at the highest possible level of detail, {\it i.e.} comparing not
only static quantities, like the $q$-dependence of the non-ergodicity
factors, but also the time dependence of the correlators in the entire
$q$ range. To our knowledge, such level of detailed 
comparison has never been
achieved, not even for the simple liquid case.
The model we propose is a 
modification of the MCT equations for simple liquids. It takes into
account the coupling between rotational and translational degrees of
freedom in a phenomenological way, by introducing a $q$-independent
coupling parameter $\chi_R$ which measures the 
increase in the strength of
the memory function $m_q(t)$ induced by the roto-translational coupling.
For $\chi_R=1$ we recover the usual MCT equations for simple liquids,
while  $\chi_R > 1$ models a system in which the rotational degrees of
freedom slow down the COM dynamics.  
In contrast to schematic models\cite{review-glass,gotze-glicerolo} 
we retain all the $q$-dependence information for the COM motion\cite{schem}. 
In our MCT model the COM dynamics in the $\alpha$-relaxation
region is entirely controlled by the value of the COM structure factor
$S_q$, by the number density $n$ and by the value of $\chi_R$.
The latter 
can be considered as a fitting parameter which controls the ideal
glass transition temperature.
The only $T$ or $P$ dependence in the dynamics
arises from the $T$ and $P$ dependence of  $S_q$ and $n$.\\
With the choice for $m_q(t)$ discussed above, 
the relaxation of the density-density correlation functions $\phi_q(t)$
in the $\alpha$-region\cite{alfa}
is described, 
in terms of a rescaled time $\hat{t} \equiv t/\tau$\cite{tau},  
by a system of coupled integro-differential equations:
\begin{equation}
\phi_q(\hat{t})=m_q(\hat{t}) -
{d \over {d \hat{t}}} \int_0^{\hat{t}} ds \,\, m_q(\hat{t}-s) {\phi}_q(s)
\label{eq:dyn}
\end{equation}
where  $m_q$  is

\begin{equation}
m_q [\phi_k(t)] = {\chi_R \over 2} \int{ {{d^3k} \over 
(2 \pi )^3} V(\vec{q},\vec{k}) \phi_k(t) \phi_{|\vec{q}-\vec{k}|}}(t) 
\label{eq:mq}
\end{equation}

\noindent
and 

\begin{eqnarray}
V(\vec{q},\vec{k}) \equiv  S_q S_k S_{|\vec{q}-\vec{k}|} {1 \over {n q^4}}~~~~~~~~~~~~~~~~~~~~~~~~~~~~ \nonumber\\
\left[ {\vec{q}} \cdot 
\vec{k}  {{(1-S_k^{-1})} } +\vec{q} \cdot  
(\vec{q}-\vec{k})  { {(1-S_{|\vec{q}-\vec{k}|}^{-1} )} }
 \right]^2
\label{eq:v}
\end{eqnarray}

\noindent
Eq.\ref{eq:dyn} and Eq.\ref{eq:v} are the usual MCT expressions while
Eq.\ref{eq:mq} differs from the one for simple liquids only by the
presence of the $q$-independent multiplicative factor $\chi_R$.
In the early $\alpha$-relaxation region 
the asymptotic analytic solution of Eq.\ref{eq:dyn} follows, to
leading order, a power law (the von Schweidler law)

\begin{equation}
\phi_q(t) \sim f_q-h_q^{(1)} \left({t \over \tau} \right)^b+h_q^{(2)}
\left({t \over \tau} \right)^{2b}+ O \left( (t/ \tau)^{3b} \right)
\label{eq:vs}
\end{equation}   

\noindent
where $f_q$, the so-called non-ergodicity factor, 
can be calculated solving 
the coupled integral equations

\begin{equation}
{f_q \over (1-f_q)}=m_q[f_k].
\label{eq:fq}
\end{equation}

\noindent
The exponent $b$ in Eq. \ref{eq:vs} 
is related to the  exponent  parameter 
$\lambda$ by

\begin{equation}
\lambda={{ \Gamma(1+b)^2 } \over { \Gamma(1+2b)}}
\label{eq:b-lambda}
\end{equation}

\noindent
where $\Gamma$ is the Euler gamma function. $\lambda$
is defined by

\begin{equation}
\!\!\!\lambda \equiv {1 \over 2} \!\! \int_0^\infty\!\!\!\!\!\!\!dq
\!\! \int_0^\infty \!\!\!\!\!\!\! dk
\!\! \int_0^\infty \!\!\!\!\!\!\! dp \,\,
%dq \,\, dk \,\, dp \,\,
\hat{e}^c_q \,(1-f_k)^2 e^c_k {{\delta^2 m_q} \over {\delta f_k \delta f_p}}
 (1-f_p)^2 e^c_p
\label{eq:lambda}
\end{equation}

\noindent
where $e^c$ and $\hat{e}^c$ are the right and left eigenvectors of the 
maximum eigenvalue of the stability matrix

\begin{equation}
C^c_{qk}[f_p] \equiv \left. {{\delta m_q[f_p]} \over {\delta f_k}}
\right|_{[f_p]} \left( 1-f_k \right)^2  
\end{equation}

\noindent
evaluated at the ideal transition temperature $T_c$.
Details on the calculation of the critical amplitudes 
$h_q^{(1)}$ and $h_q^{(2)}$ can be found in Ref.\cite{franosch}.\\
In the late $\alpha$-region it is not possible to
solve analytically Eq.\ref{eq:dyn}, but the numerical solution 
is well fitted  by a stretched exponential law 
(Kohlrausch-William-Watts law)

\begin{equation}
\phi_q(t) \sim A^K_q \exp \left[ {-{ \left( {t \over 
{\tau^K_q}} \right)^{\beta^K_q}}} \right] 
\label{eq:KWW}
\end{equation}

\noindent 
where $\beta^K_q \le 1$.\\
We compare the predictions of the semi-schematic model
(Eqs.\ref{eq:dyn}-\ref{eq:v}) with the corresponding correlators
evaluated in a long MD simulation\cite{sgtc}. 
The simulation was performed using the SPC/E 
potential\cite{spce}, a model developed to simulate
the behavior of liquid water. It 
treats the molecule as a rigid unit
and models the pair interactions as a sum of electrostatic and
Lennard Jones terms. Previous studies have shown that  SPC/E
reproduces semi-quantitatively the static and dynamic properties of
liquid water\cite{spce}. In water,   an open-structured
liquid with an average coordination number around 4,
the slowing down of the dynamics on supercooling is {\it not}
related to packing constraints but it is driven by the formation of a
tetrahedral network of strong and highly directional
bonds. 
The role of the
cages in reducing the mobility of the molecules is played by the
hydrogen bonds which freeze the rotational motion of the 
molecules and therefore hinder their translational motion.  The
steric cages which slow down the dynamics in simple liquids
are replaced by ``energetic'' cages.\\
It has been shown numerically that the
SPC/E dynamics in supercooled states is
characterized by a two step relaxation process, 
by a self-similar region
in the early $\alpha$-relaxation region 
(with a von-Schweidler exponent $b=0.50 \pm 0.05 $)
and by a stretched exponential decay in the long time $\alpha$ regime.
The relaxation times display a power law divergence at the 
 critical
temperature $T_c =(200 \pm 2) K$, which 
 has been identified with the ideal MCT
transition temperature\cite{sgtc}. Moreover, 
the rotational correlators follow closely the COM correlators,
supporting the view that hydrogen bonds form strong energetic
cages and that molecules can not rotate within a cage. The process of
breaking and reforming cages, characteristic of the
$\alpha$-relaxation region is the bottleneck for both COM and
rotational diffusion.\\ 
We solve numerically Eqs.\ref{eq:dyn}-\ref{eq:v} on a grid of 300
equispaced $q$ vectors, using as input the COM number density
($n=32.3\, nm^{-3}$) and the COM $S_q$ evaluated from the MD data.  We
study the model for different values of the coupling $\chi_R$.  When
$\chi_R= 1.93$, the ideal glass transition temperature in the model
coincides with the $T_c$ detected in SPC/E water\cite{sgtc}.  Since
with this choice of $\chi_R$ the ideal glass transition temperatures
in the model and in SPC/E are the same, we can compare, without using
further fitting parameters, results from the MD simulations and
theoretical predictions for all $q$-vectors.\\
The comparison\cite{fit} for the $q$-dependence of $f_{q}$, $h_q^{(1)}$ and
$h_q^{(2)}$ in Eq.\ref{eq:vs} is shown in Fig. \ref{fig:fq-h1-h2}
\cite{scaling}.  For reference, we also show $S_q$, the only
$q$-dependent input in the theory. As in the simple liquid case, the
static quantities show oscillations in phase with $S_q$.  We stress,
in passing, the peculiar shape of the COM structure factor $S_q$ for
SPC/E water when compared to the $S_q$ of simple liquids.  The
presence of the pre-peak around $q=18 nm^{-1}$ (the analog of the
so-called first sharp diffraction peak in silica) is the hallmark of
medium range order characteristic of network-forming
liquids\cite{elliot,foley}.\\
The agreement
between numerical and theoretical values for $f_{q}$ and $h_q^{(1)}$ is
striking although deviations at large $q$ are clearly seen; the
difference between theory and simulations for large values of $q$
({\it i.e.} small distances) is not surprising since this is the
region where the system is more sensitive to the detailed geometry of
the molecules, completely neglected by the model.  The agreement for
$h_q^{(2)}$ is not as good.
We recall that  the numerical values for
$h^{(2)}_q$ were obtained by a fitting of $\phi_q(t)$ 
to a second order polynomial in
$(t/\tau)^b$ (Eq.\ref{eq:vs}) and thus the fitted values
are affected by higher
order corrections; on the contrary the theoretical $h^{(2)}_q$ is the
exact coefficient of the second order term $(t/\tau)^{2b}$.\\ 
From the calculated $f_q$ values, we evaluate the
exponent parameter $\lambda$
by numerical integration of Eq. \ref{eq:lambda}. We  obtain 
$\lambda=0.79$ which yields, through 
Eq. \ref{eq:b-lambda}, $b=0.49$. The value for the exponent  $b$,
which controls the self-similar dynamics in the early $\alpha$-region,
is in excellent agreement with the one calculated by
MD, {\it i.e.} $b=0.50 \pm 0.05$. Thus,
the  discrepancies between theory and numerical data 
at large $q$-values
do not affect  the value of the von Schweidler exponent $b$.
Indeed, the leading 
contributions to the integral in Eq.\ref{eq:lambda} come
from the maxima of $S_q$, {\it i.e.} from the values at the pre-peak  
and at the main peak, where our model
provides the correct results.\\
To extend the comparison to the late 
$\alpha$-region, we solve Eq.\ref{eq:dyn} for all 
300 $q$-vectors\cite{algorithm} and 
we fit, for each $q$,  
the time evolution of the MD and theoretical correlators 
to a Kohlrausch stretched exponential according to
Eq.\ref{eq:KWW}.
The $q$-dependence of the fit parameters  (amplitude $A^K_q$, 
relaxation time $\tau^K_q$ and stretching exponent $\beta^K_q$)
is compared with the analogous quantities obtained from the MD data 
in Fig.\ref{fig:stretched}.  Again, we
find a remarkable agreement 
between the theoretical predictions and the MD
data in the low $q$ region, both in the 
amplitudes and in the relaxation times. Instead, the
parameters $\beta^K_q$ show oscillations in phase with $S_q$
but the theoretical and MD values  
differ up to 25\%. This indicates
that in the MD data,  the $\alpha$-relaxation dynamics covers
a larger time range. 
The differences in  $\beta^K_q$  can be
tentatively associated with the 
drastic reduction in the number 
of coupled correlators retained in the present 
model, which neglects the angular correlators.\\
In summary, a simple modification of the
MCT equations for simple liquids is able to describe 
the $\alpha$-relaxation process in SPC/E water, a
model for a molecular network-forming  
liquid with strong directional bonds. 
By using as input the fixed value of $n$ and the
COM structure factor $S_q$, the theory is able to 
describe both the $q$- and $t$-dependence of the
density-density correlations in a satisfactory way. 
We stress again that
except for the value of $\chi_R$ and for a $q$-independent scaling of time
relating $\hat t$ with real time\cite{scaling}, no fitting parameters are
involved in the comparison.
The model can be extended to
describe the self-particle dynamics and the collective transverse 
currents\cite{next}.\\
The approach presented here is a first step 
towards a full MCT description, which will necessarily include 
orientational correlations, {\it i.e.} the geometry of the molecule
\cite{schilling,theis,franosch-dumbell,kawasaki}.
Our semi-schematic approach 
provides evidence that the slow dynamics in complex
molecular liquids can be {\it quantitatively} described by 
a very simple modification of the MCT equations for
simple liquids, thus retaining the simplicity of the COM description.

%\end{multicols}
        
%\newpage

\begin{figure}
%\centerline{\psfig{figure=h1h2-semi-schem.eps,height=15cm,width=20cm,clip=,angle=0.}}
\caption{Comparison between theory (solid lines) and MD simulations (symbols)
for the non-ergodicity parameter $f_q$ and the critical amplitudes 
$h^{(1)}_q$ and $h^{(2)}_q$ in Eq.\protect\ref{eq:vs}. 
(See Ref.\protect\cite{scaling}).
The center of mass structure 
factor $S_q$ is shown as a reference. MD data are from Ref.
\protect\cite{sgtc}. } 
\label{fig:fq-h1-h2}
\end{figure}

%\newpage

\begin{figure}
%\centerline{\psfig{figure=stretch-semischem.eps,height=15cm,width=20cm,clip=,angle=0.}}
\caption{Fitting parameters to the Kohlrausch-William-Watts 
law (Eq.\protect\ref{eq:KWW}) as obtained by the solution of 
Eq.\protect\ref{eq:dyn}
(solid lines) and by the MD data (symbols). The time scale for the 
MD values of $\tau^K_q$ is $1$ {\it ps}. In the theoretical values 
of $\tau^K_q$ the time scale has been chosen in an appropriate way
as explained in Ref.\protect\cite{scaling}.
MD data are from Ref.\protect\cite{sgtc}. 
} 
\label{fig:stretched}
\end{figure}

\setcounter{figure}{0}

\eject

\begin{figure}
\centerline{\psfig{figure=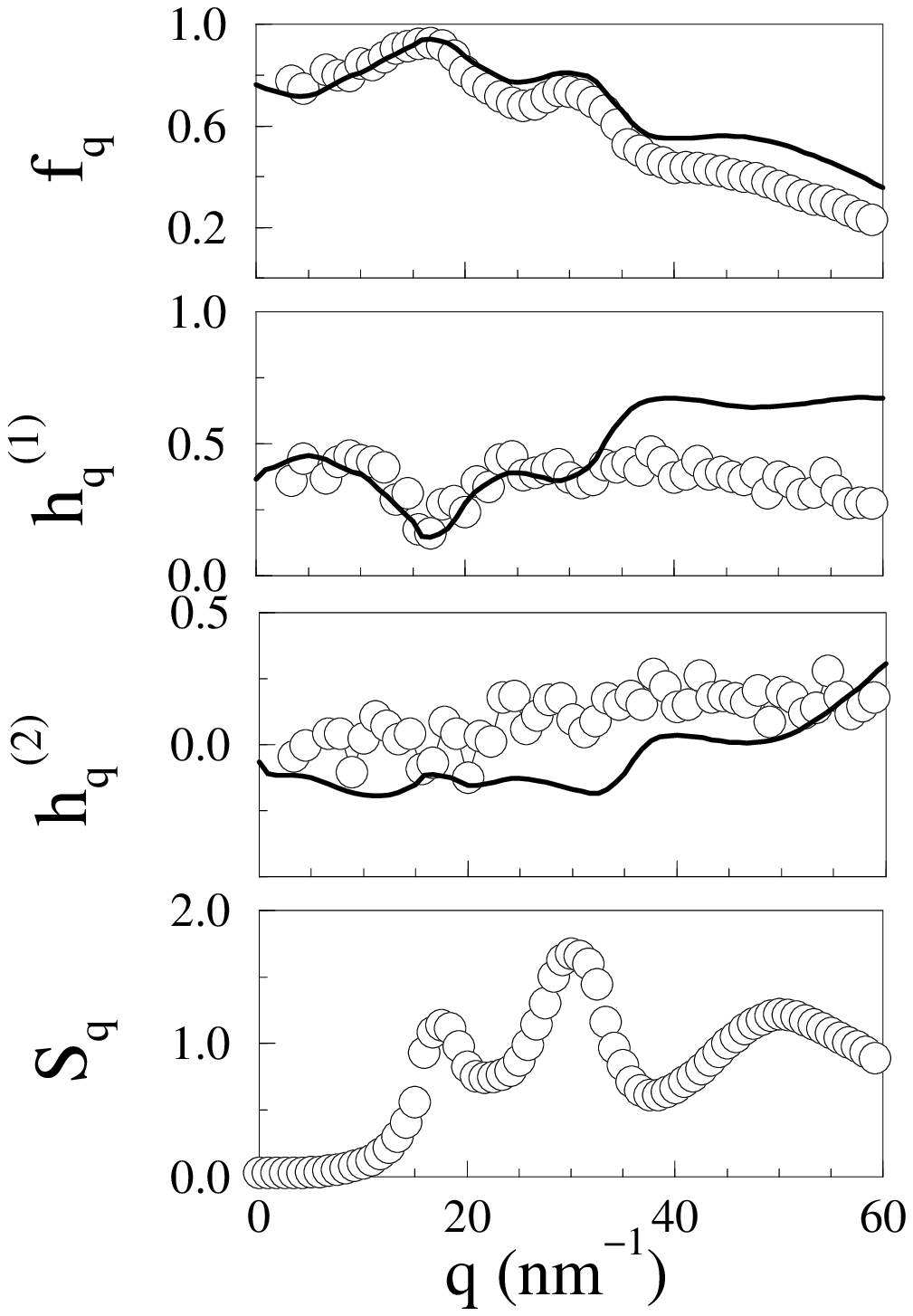,height=20cm,width=26cm,clip=,angle=0. }}
\caption{L. Fabbian et al.}
\end{figure}

\eject

\begin{figure}
\centerline{\psfig{figure=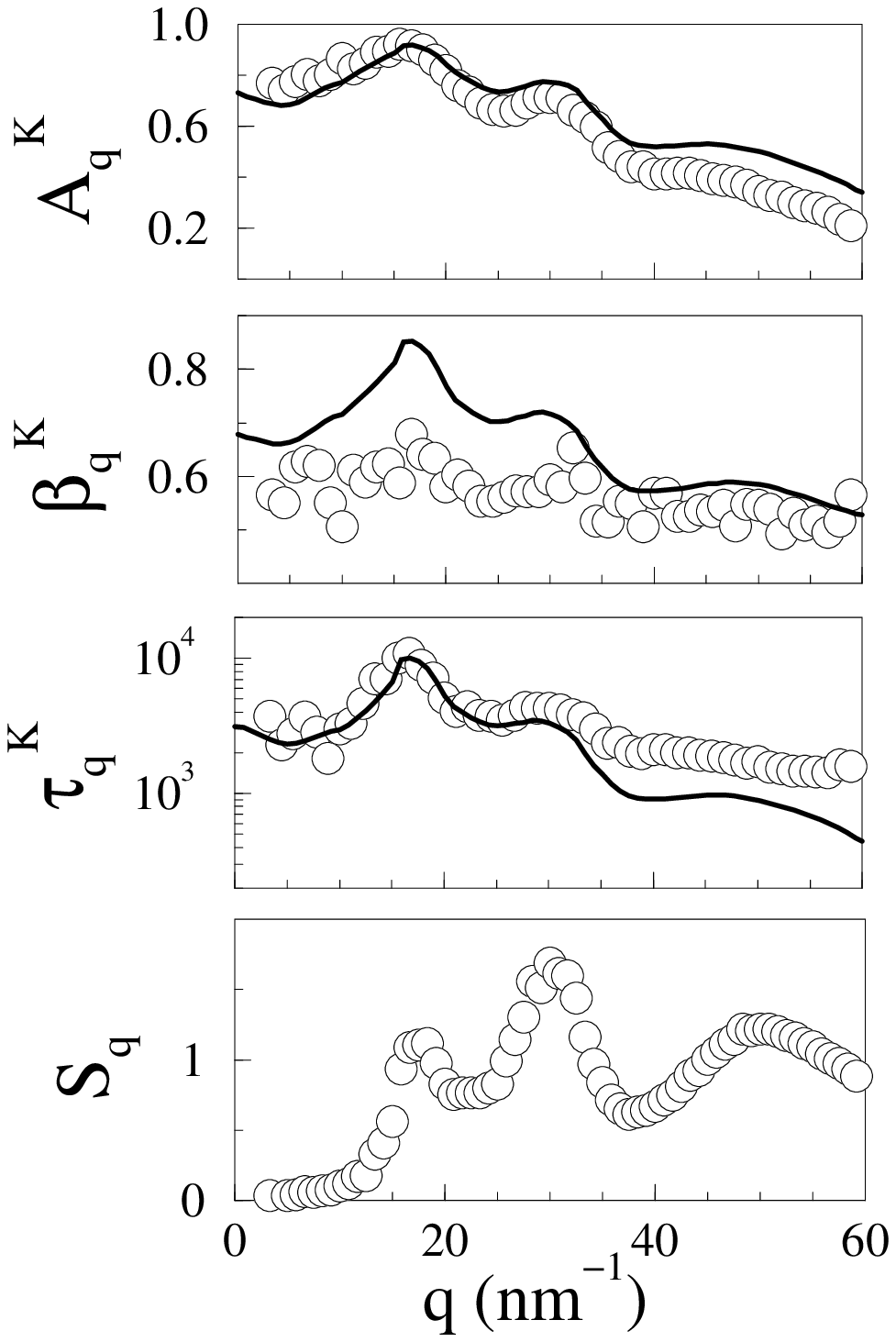,height=20cm,width=26cm,clip=,angle=0. }}
\caption{L. Fabbian et al.}
\end{figure}


\begin{references}

\bibitem{review} For recent reviews see for example 
C.A. Angell, Proc. 1996 Enrico Fermi Summer School in Physics, 
Italian Physical Society, in press.;
R. Schilling in {\it Disorder Effects on Relaxation Processes}, 
Eds. R. Richert and A. Blumen (Springer, Berlin 1994)

\bibitem{review-glass} W. G\"{o}tze, 
in {\it Liquids, Freezing and the Glass Transition}, 
Eds.:J. P. Hansen, D. Levesque and J. Zinn-Justin, Les Houches. 
Session LI, 1989, (North-Holland, Amsterdam, 1991).

\bibitem{cummins} G. Li, W.M. Du, X.K. Chen and H.Z. Cummins {\em
Phys. Rev. A} {\bf 45}, 3867 (1992).  H.Z. Cummins {\it et al.}  {\em
Phys. Rev. E} {\bf 47} 4223 (1993).  N.J. Tao {\it et al.} {\em
Phys. Rev. A} {\bf 44}, 6665 (1991).


\bibitem{andalo} See for example {\em Phil. Mag.}, proceedings
of the 6th International Workshop on Disordered Systems (Andalo 1997),
in press.

\bibitem{barrat}
J.-L. Barrat and A. Latz, {\em J.Phys.: Condens.Matter} {\bf 2} 4289-4295
(1990).

\bibitem{walstrom}  L.J. Lewis and  G. Wahnstrom, {\em Phys. Rev. E}  
{\bf50},  3865 (1994).

\bibitem{kob} W. Kob and H.C. Andersen, {\em Phys. Rev. E} {\bf
 51}, 4626 (1995) and {\em Phys. Rev. E} {\bf52}, 4134 (1995).
M. Nauroth and W. Kob, {\em Phys. Rev. E} {\bf55}, 675 (1997).


\bibitem{Gotze} W. G\"otze and L. Sj\"ogren, {\em Rep. Prog. Phys.} {\bf
55}, 241 (1992).

\bibitem{extended}
Extension of the {\it ideal} MCT to the case where current density
fluctuations are included among the slow modes generates the so-called
{\it extended} MCT, in which phonon assisted hopping processes are
also considered as candidate for structural relaxation processes.
These processes smear out the ideal liquid-glass transition.
See for example W. G\"otze and A. Sj\"ogren,  {\em Transport Theory
and Statistical Physics} {\bf 24}, 801 (1995).


\bibitem{parisi}
S. Franz and G. Parisi, {\em J. Phys. I} (France) {\bf 5} 1401 (1995).

\bibitem{alba}
C. Alba-Simionesco and M. Krauzman, {\em J. Chem. Phys.},  {\bf 102} 
6574 (1995).

\bibitem{sokolov} 
A. P. Sokolov, J. Hurst and D. Quitmann, {\em
Phys. Rev. B} {\bf 51}, 12865 (1995).

\bibitem{kob-dumb} S. K\"ammerer, W. Kob and R. Schilling,
preprint (1997).

\bibitem{sgtc} F. Sciortino, P. Gallo, P. Tartaglia, S.-H. Chen, {\em
Phys. Rev. E} {\bf 54} 6331 (1996). F. Sciortino, L. Fabbian, S.H. Chen
and P. Tartaglia, submitted to {\em Phys. Rev. E} (1997).

\bibitem{gotze-glicerolo} T. Franosch, W. G\"{o}tze, M.R. Mayr and A.P. Singh,
{\em Phys. Rev. E}  {\bf55}, 3183 (1997).

\bibitem{tei} H. Teichler, {\em Phys. Rev. Lett.} {\bf 76} 62 (1996).

\bibitem{schilling}
R. Schilling and T. Scheidsteger, preprint. See also
{\em Phil. Mag.} (1997) in press. 

\bibitem{theis} 
C. Theis, Diplom Thesis Johannes Gutenberg Universit\"at (1997).

\bibitem{franosch-dumbell} T. Franosch, M. Fuchs, W. G\"otze, M.R. Mayr and 
A.P. Singh, preprint.

\bibitem{kawasaki} K. Kawasaki, {\em Physica A}, preprint (1997).

\bibitem{schem} The ideal MCT equations reduce to the schematic model
with memory function $m(t) \propto \phi_0^2(t)$ if the $q$-dependence 
of the correlators is 
condensed in a single representative $q_0$-vector, {\it i.e.} if
the structure factor is $S_q \propto \delta (q-q_0)$ (see E. Leutheusser, 
{\em Phys. Rev. A}, {\bf 29}, 2765 (1984)). We call our model 
{\it semi-schematic} because we retain all the $q$-dependence of the 
translational correlators but we model the coupling with the angular 
correlation functions with a single $q$-independent value $\chi_R$. 


\bibitem{alfa} The time evolution of all correlators in the supercooled regime
follows a two-step decay. After the short time microscopic dynamics, 
the correlators approach a plateau value $f_q$. 
The decay after the plateau value is
commonly referred to as $\alpha$-relaxation.

\bibitem{tau} According to MCT in the $\alpha$-relaxation region
a characteristic $q$-dependent time scale $\tau$ appears, 
which strongly depends on
the temperature through the scaling relation $\tau \propto |T-T_c|^{-\gamma}$.
The divergence of $\tau$ defines the {\it ideal} critical temperature $T_c$.

\bibitem{franosch} T. Franosch, M. Fuchs, W. G\"otze, M.R. Mayr and 
A.P. Singh, {\em Phys. Rev. E} in press.

\bibitem{spce} H.J.C. Berendsen, J.R. Grigera and T.P. Straatsma,
{\it J. Phys. Chem.} {\bf 91}, 6269 (1987).

\bibitem{fit}
We recall that the MD non-ergodicity parameter and the critical amplitudes
are obtained fitting the calculated time evolution of the density correlator 
in the early $\alpha$ region using the von Schweidler's 
law (\protect\ref{eq:vs}). (See Ref. \protect\cite{sgtc}).

\bibitem{scaling} Due to the scale invariance of 
Eq.\protect\ref{eq:dyn} the values for $\tau$ are
defined up to a $q$-independent multiplicative number which is a
fitting parameter in the comparison between theory and simulations
for $h^{(1)}_q$, $h^{(2)}_q$ and $\tau^K_q$. 

\bibitem{elliot} A. Uhlherr and S.R. Elliott   {\em Phil. Mag. B}
{\bf 71}, 611 (1995).

\bibitem{foley} Foley {\it et al.}, {\em Phil. Mag. B}, 557 (1995).

\bibitem{algorithm} We use the numerical algorithm
described in A.P. Singh, Ph.D. Thesis, Technischen 
Universit\"at M\"unchen, (1997). See also  
W. G\"otze {\em J. Stat. Phys.} {\bf 83} 1183 (1996).

\bibitem{next} L. Fabbian, F. Sciortino, F. Thiery and P. Tartaglia 
(in preparation).

\end{references}
\end{document}